\documentclass[10pt, twocolumns]{IEEEtran}
 \makeatletter

\newcommand{\Rmnum}[1]{\expandafter\@slowromancap\romannumeral #1@}
\makeatother
\usepackage{graphicx,cite,epsfig,amssymb,amsmath,multirow}
\usepackage{graphicx,amsmath,amssymb,amsfonts,cite}
\usepackage{mathrsfs}
\usepackage{algorithm}
\usepackage{algorithmic}
\usepackage{float}
\usepackage{bm}
\usepackage{stfloats}
\usepackage{ulem}
\usepackage{color}

\makeatother
\newcommand{\bs}{{\bf s}}
\newcommand{\ba}{{\bf a}}
\newcommand{\bG}{{\bf G}}
\newcommand{\bA}{{\bf A}}
\newcommand{\bB}{{\bf B}}
\begin{document}
\bibliographystyle{ieeetr}

%------------------------------- title ----------------------------------------
\title{Improving Bandwidth Efficiency of FBMC-OQAM Through Virtual Symbols}

\author{\normalsize
Daiming Qu, Fang Wang, Tao Jiang, \textit{Senior Member}, \textit{IEEE}, and Behrouz Farhang-Boroujeny, \textit{Senior Member}, \textit{IEEE}
\thanks{
Copyright (c) 2013 IEEE. Personal use of this material is permitted.
However, permission to use this material for any other purposes must
be obtained from the IEEE by sending a request to
pubs-permissions@ieee.org.

Daiming Qu, Fang Wang, Tao Jiang are with the
School of Electronic Information and Communications, Huazhong University of
Science and Technology, Wuhan, 430074, China (e-mail: wf201371743@gmail.com; qudaiming@hust.edu.cn; Tao.Jiang@ieee.org).

Behrouz Farhang-Boroujeny is with the Electrical and Computer Engineering Department,
University of Utah, Salt Lake City, UT 84112 USA (e-mail: farhang@ece.utah.edu).

}\\}
\markboth{xxxxxxxxxxx} {{et al.}: Time Domain Channel Estimations for OQAM-OFDM
Systems}
\maketitle

\begin{abstract}
Filter bank multicarrier (FBMC) systems that are based on offset quadrature amplitude modulation (OQAM), namely, FBMC-OQAM, have been criticized for their inefficiency in the use of spectral resources, because of the long ramp-up and ramp-down tails at the beginning and the end of each data packet, respectively.
We propose a novel method for shortening these tails. By appending a set of virtual (i.e., none data carrying) symbols to the beginning and the end of each packet, and clever selection of these symbols, we show that the ramp-up and ramp-down tails in FMBC-OQAM can be suppressed   to an extent that they deem as negligible and thus may be ignored. This shortens the length of signal burst in each FBMC-OQAM packet, hence, improves on its bandwidth efficiency, viz., the same data is transmitted over a shorter period of time. We develop an optimization method that allows computation of virtual symbols, for each data packet. Simulation results show that, compared to existing methods,  the proposed tail-shortening approach leads to a superior out-of-band (OOB) emission performance and a much lower error vector magnitude (EVM) for the demodulated symbols.

\begin{keywords}
virtual symbols, FBMC-OQAM, EVM, spectral efficiency.
\end{keywords}
\end{abstract}
%----------------------------- I. INTRODUCTION -----------------------------------
\section{Introduction}\label{s1}
In the past, orthogonal frequency division multiplexing (OFDM) has enjoyed its dominance as the most popular signaling method for broadband communications, \cite{PrasadBook}, \cite{LiStuberBook}. Most of the broadband standards, as of today, have adopted OFDM. OFDM offers a minimum complexity, achieves a fair bandwidth efficiency, and possibly is the best compromise choice for point-to-point communication.   However, it has been noted that OFDM has to face many challenges when considered for adoption in the more complex networks. For instance, the use of OFDM in the uplink of multiuser networks, known as OFDMA (orthogonal frequency division multiple access), requires accurate synchronization of the users' signals at the base station input, \cite{Morelli2007}. Such synchronization is not straightforward and demands a lot of resources.

Taking note of these issues, currently, many researcher are exploring alternative waveforms for the next generation of wireless communications; the 5G. Researchers in the Alcatel-Lucent Bell Laboratories have proposed and studied a modified OFDM waveform that they call universal filtered multicarrier (UFMC), \cite{UFMC1,UFMC2,UFMC3,UFMC4,UFMC5}. In UFMC, each bundle of adjacent subcarriers that belong to a user are passed through a filter to minimize multi-user interference, equivalently, multiple access interference (MAI). To allow good filtering, while its bandwidth efficiency is kept at the same level as OFDM, in UFMC, there is no cyclic prefix (CP) and, instead, the CP interval is reserved to absorb the transient of the underlying filters.

Generalize frequency division multiplexing (GFDM) is another popular waveform that has been proposed by Fettweis et al., \cite{GFDM2009,GFDMinvited2014}. GFDM may be thought of as a modified OFDM waveform, where each subcarrier is shaped by a high quality filter. Moreover, to keep the spectral efficiency high, GFDM is designed such that one CP is added to a block of $N$ data symbols that are spread across $K$ subcarriers and $M=N/K$ time slots.  Also, to allow the addition of CP, the subcarriers filtering operation in GFDM is based on a circular convolution (not a linear convolution).

The third contender for 5G builds its waveform based on filter bank multicarrier with offset quadrature amplitude modulation (FBMC-OQAM) at the same time filtering operations are based on circular convolution, as in GFDM, \cite{Jianglei,C-FBMC2014,C-FBMCinvited2014,C-FBMC2014a}. It is thus named cyclic FBMC (C-FBMC). Also, similar to GFDM, a CP is added to each block of data symbols. In this paper, for brevity, we remove the suffix OQAM from FBMC. Hence, any mention of FBMC should be understood  as being FBMC with OQAM signaling.
%that has also recently appeared in the literature and builds based on the same concept as GFDM, is the cyclic filter bank multicarrier (C-FBMC), \cite{references}. The filtering operation in C-FBMC follows that of ,   \cite{Farhang-Boroujeny2011}, and adopts circular convolution to allow use of a CP per data block.

The presence of CP in GFDM and C-FBMC allows multi-users in the uplink of a network to be quasi-synchronous, hence, perfect synchronization can be relaxed. In UFMC, this relaxed synchronization is allowed because of the presence of soft ramp-up and ramp-down at the two sides of each symbol. However, one should note that still GFDM, C-FBMC, and UFMC require synchronization of the mobile nodes to a great degree.

Literatures on UFMC, GFDM, and C-FBMC, beside reviewing the shortcomings of OFDM for 5G, have also criticized the conventional FBMC, \cite{Farhang-Boroujeny2011, Siohan2002, Gao2011, Hirosaki1981, Chen2013, Zhang2012, Ihalainen2011}, where filtering is based on linear convolution. It has been noted that such waveform are bandwidth inefficient because of the signal ramp-up and ramp-down (arising from filtering transients) at the beginning and the end of each data packet, respectively. However, we note that FBMC has an advantage over  UFMC, GFDM, and C-FBMC as users can be completely asynchronous. Taking note of this important property of FBMC, this paper revisits FBMC with the goal of redesigning the waveform such that  it will have very short ramp-up and ramp-down, hence, bringing FBMC bandwidth efficiency to a level comparable to those of UFMC, GFDM, and C-FBMC.

Another point that needs our attention here is the following. In GFDM and C-FBMC literature, it has often been emphasized that each data packet makes use of only one CP, and thus argued that these waveforms offer a very high bandwidth efficiency. This may not remain true in some practical cases, where long packets of data have to be transmitted and hence channel variation over each packet interval may not remain negligible. In such cases, each packet should be divided into a number of shorter blocks and GFDM/C-FBMC  waveform synthesis, including addition of CP, is applied to each block. Pilot symbols are also added into each block to assist the receiver in tracking channel variations, following the methods that have been developed for OFDM, \cite{coleri2002channel}. Noting this and the fact that FBMC packets for any size have only one ramp-up and one ramp-down implies that FBMC for long packets offers a higher bandwidth efficiency than GFDM, C-FBMC, and UFMC.

In the past, couple of trivial methods have been proposed to reduce ramp-up and ramp-down durations in FBMC. These methods have been referred to by the generic name of {\em tail-shortening}. The first tail-shortening approach aplies a hard truncation to the signal tails, \cite{M.Bellanger}. Although this method totally removes the tail, the truncation causes intersymbol interference (ISI) as well as intercarrier interference (ICI) on the data symbols near the edges of the data packet. In addition, it increases the out-of-band (OOB) emission due to sharp truncation of the signal. An improved method,  \cite{Ihalainen}, uses a generalized weighting window to smoothen the edge transitions introduced by truncation. This  improves the OOB emission performance, however, it introduces more ISI/ICI than the hard truncation method.

In this paper, we propose a more intelligent method that avoids ISI/ICI and keeps OOB emission similar to that of the FBMC without tail-shortening. We introduce additional, but redundant, symbols at the two sides of the packet and these symbols are chosen such that to suppress the waveform tails without introducing any ISI/ICI and without increasing OOB emission. Since these symbols do not carry any data, we refer to them as {\em virtual symbols}.

The rest of this paper is organized as follows. Section~\ref{sec_model} presents a brief description of FBMC-OQAM signaling method. Section~\ref{sec_review} gives a review of the tail truncation methods in the past literature. The proposed tail-shortening method by virtual symbols is presented in Section~\ref{sec_virtual}. A few considerations related to the receiver implementation are discussed in Section~\ref{sec_considerations}. The performance of the proposed method is compared with other tail-shortening methods  through computer simulations in Section~\ref{sec_perf}. The concluding remarks are made in Section~\ref{sec_conlusion}.

\section{FBMC-OQAM System Model}\label{sec_model}

\begin{figure}
\centering
\includegraphics[scale=0.8]{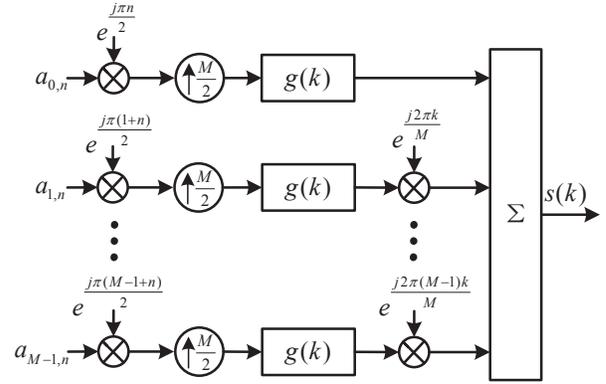}
\caption{Discrete time domain model of an FBMC-OQAM transmitter.}
\label{fig1}
\end{figure}

The discrete time domain model of an FBMC-OQAM transmitter is shown in Fig. \ref{fig1}. The subcarrier spacing is $1/T$, where $T$ is the time interval between QAM symbols. Each complex-valued symbol is partitioned into a pair of real-valued/pulse amplitude modulated (PAM) symbols. The PAM symbol at the time-frequency index $(m,n)$ is denoted by $a_{m,n}$, where $m$ is the frequency index and $n$ is the time index. Moreover, $a_{m,2\ell}$ and $a_{m,2\ell+1}$ are the real and imaginary parts of the $\ell$-th OQAM symbol, respectively. They  are $T/2$ spaced. We assume that there are $M$ samples per each symbol interval.

Assuming that $s(k)$  is a burst of length $N$ of PAM symbols, following Fig. \ref{fig1}, one finds that
\begin{equation}\label{eq 1}
s(k) = \sum\limits_{m \in \Omega } {\sum\limits_{n = 0}^{N - 1} {{a_{m,n}}g\left({k - n\frac{M}{2}}\right){e^{j2\pi mk/M}}} {e^{j(m + n)\pi /2}}},
\end{equation}
where $\Omega$ is the set of active subcarriers and $g(k)$ is the impulse response of the prototype filter. The prototype filter has a length of $L_{\rm filter}=\eta M+1$, where $\eta$ is the overlapping factor (i.e., the number of FBMC symbols that overlap in time). Accordingly, the burst length $L_{\rm burst}$ is equal to $(N-1)M/2+\eta M$.

\section{Review of Tail-Shortening Methods}\label{sec_review}

Since FBMC-OQAM systems adopt filters with rather long length, the resulting ramp-up at the beginning and ramp-down at the end of each data burst cover multiple symbol intervals. This considerably reduces the gain of FBMC-OQAM in spectral efficiency, when data bursts are short.
In what follows, we use the tail at the end of the burst to illustrate the tail-shortening methods. The processing at the start of the burst is similar.

The most trivial method of shortening the bust length is hard truncation of the tails. For the end part of the burst, this is  expressed as
\begin{equation}\label{eq 2}
{s_{\rm h.trunc}}(k) = \left\{ \begin{array}{ll}
s(k),&k \le {K_{\rm e.burst}}\\
0,&k > {K_{\rm e.burst}}
\end{array} \right.,
\end{equation}
where $K_{\rm e.burst}$ is the end of the data burst after truncation, and subscript ``h.trunc'' on ${s_{\rm h.trunc}}(k)$ is to emphasize it is hard truncated.
The cost of this method is an increased OOB emission and ISI/ICI due to the truncation.

Another method, called truncation with windowing, uses a generalized weighting window to shorten the tail. The windowed signal is given as
\begin{equation}\label{eq 3}
{s_{\rm w.trunc}}(k) = s(k)w(k),
\end{equation}
where $w(k)$ is the window function. Specifically, we consider a raised-cosine window, which is expressed as
\begin{equation}\label{eq 4}
w(k) =
\begin{cases}
1,& k < K_{\rm b.ro}\\
\frac{1}{2} + \frac{1}{2}\cos \left( {\frac{{(k - K_{\rm b.ro})\pi }}{{{L_{\rm ro}}}}} \right),& K_{\rm b.ro} \le k \le {K_{\rm e.burst}},\\
0,& k > {K_{\rm e.burst}}
\end{cases}
\end{equation}
where $L_{\rm ro}$ is the length of roll-off and $K_{\rm b.ro}=K_{\rm e.burst} - L_{\rm ro}$ is the beginning of the roll-off.
Although this method smoothens the edge transitions introduced by truncation, hence, reduces the OOB emission, it results is a greater ISI/ICI than the hard truncation method. This is because the roll-off introduces an additional distortion to the signal.

Furthermore, as discussed earlier, the rump-up and ramp-down issue may be resolved by replacing the linear convolution in  FBMC by a circular one and use a CP to take care of channel transient. This leads to C-FBMC signaling, \cite{Jianglei,C-FBMC2014,C-FBMCinvited2014,C-FBMC2014a}. In addition, to avoid OOB emission caused by hard truncations at the beginning and end of each bust, a soft truncation with windowing, similar to (\ref{eq 3}), may also be applied. Such windowing to be as effective as FBMC, particularly for frequency points near the passband edges of transmission can increase the length of C-FBMC to that of FBMC  (with linear convolution). %Numerical examples confirming this fact is presented in Section~\ref{sec_perf}.

%A variation of truncation with windowing is the so-called circular convolution with windowing \cite{Jianglei}, which is C-FBMC with windowed truncation. This method has very low ISI/ICI due to the fact that the original FBMC signal can be recovered from a single period of the circular convolved signal and the signal affected by window does not participate in the detection at the receiver. However, since the windowing is applied on the full-power C-FBMC signal, rather than on the transition portion of the signal as the windowing on FBMC signal, it causes higher OOB emission than the windowing of FBMC signal.

\section{Tail-Shortening by Virtual Symbols}\label{sec_virtual}

\begin{figure}
\centering
\includegraphics[scale=0.6]{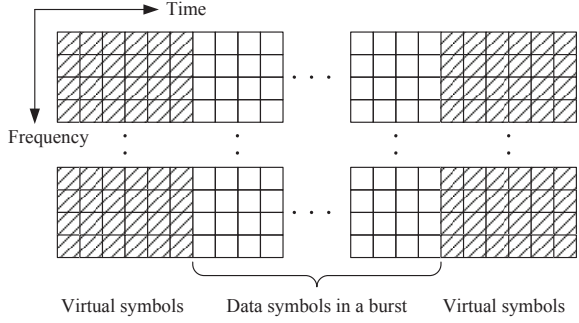}
\caption{Virtual symbols in an FBMC burst.}
\label{fig__virtual_symbols}
\end{figure}

In this paper, we propose adding a few {\em dummy symbols} at the two sides of each data burst, as demonstrated in Fig. \ref{fig__virtual_symbols}, to suppress the tails of $s(k)$. After suppressing the tails, the residual tails may be truncated with a minimum effect on OOB and/or ISI/ICI interference. Alternatively, the suppressed tails may be allowed to overlap among successive data packets. Such overlap, as our numerical results show, introduces a negligible level of interference.

We refer to the dummy symbols as {\em virtual symbols}. One may think of virtual symbols as a means of generating a cancellation signal at the tails of each FBMC burst, hence, removing the tails of the bursts.   Moreover, since this signal is synthesized following the FBMC-OQAM construction, it suppresses the tails without introducing any OOB emission as well as no ISI/ICI.

Let $\bf{A}$, with the elements of $A(p,q)$, denote the matrix of virtual symbols at the end of the burst. This results in the cancelation signal
\begin{equation}\label{eq cancelation_signal}
s_{\rm vs}(k) = \sum\limits_{p \in \Omega} {\sum\limits_{q = N}^{N + V - 1} {A(p,q)g({k - q\frac{M}{2}}){e^{j2\pi pk/M}}{e^{j(p + q)\pi /2}}} }.
\end{equation}
where $V$ is the number of virtual symbols along the time. With the addition of $s_{\rm vs}(k)$, we get the tail-shortened/suppressed signal
\begin{equation}\label{eq 9}
{s_{\rm vs.short}}(k) = s(k) + {s_{\rm vs}}(k)
\end{equation}
where `vs' in the subscript is to emphasize that virtual symbols are used for tail-shortening.

To find the virtual symbols, i.e., the elements of $\bf A$, we define the cost functions
\begin{equation}\label{eqn:xi1}
 \xi_1=   \sum_{k>K_{\rm e.burst}}|s_{\rm vs.short}(k)|^2
\end{equation}
and
\begin{equation}\label{eqn:xi2}
 \xi_2=   \sum_{k\le K_{\rm e.burst}}|s_{\rm vs}(k)|^2
\end{equation}
and search for a choice of $\bf A$ that minimizes $\xi_1$ and $\xi_2$ jointly.
Minimization of $\xi_1$ suppresses the tail of the burst, beyond $K_{\rm e.burst}$. Minimization of $\xi_2$, on the other hand, is to assure that the total energy of $s_{\rm vs}(k)$ over the range of $k\le K_{\rm e.burst}$ is kept minimal. This is necessary to avoid an unexpected increase of $|s_{\rm vs.short}(k)|$ over this range, hence, avoid an undesirable increase of peak-to-average power ratio (PAPR) of the burst.

To obtain a balanced minimization $\xi_1$ and $\xi_2$, we find $\bf A$ that minimizes the combined cost function
\begin{equation}\label{eqn:xi}
\xi=\xi_1+\gamma\xi_2
\end{equation}
where $\gamma$ is a positive parameter that should be found empirically. To derive a convenient formulation for this minimization problem, we proceed as follows.

We define the pair of column vectors $\bs_1$ and $\bs_2$ consisting of elements $\{s_{\rm vs.short}(k),\mbox{ for } k>K_{\rm e.burst}\}$ and $\{s_{\rm vs}(k),\mbox{ for } k\le K_{\rm e.burst}\}$, respectively. We also note that $\bs_1$ and $\bs_2$ can be expanded as
\begin{equation}\label{eqn:s1}
\bs_1=\bs+\bG_1\ba
\end{equation}
and
\begin{equation}\label{eqn:s2}
\bs_2=\bG_2\ba
\end{equation}
where $\bs$ contains samples of $s(k)$, for $k>K_{\rm e.burst}$, $\ba$ is  constructed by rearranging the columns of $\bA$ in a column vector, and the matrices $\bG_1$ and $\bG_2$  are obtained trivially by taking note of the relationship (\ref{eq cancelation_signal}).

Using (\ref{eqn:s1}) and  (\ref{eqn:s2}) in  (\ref{eqn:xi1}) and  (\ref{eqn:xi2}), respectively, and the results in  (\ref{eqn:xi}), we obtain
\begin{equation}\label{eqn:xi2}
\xi=\|\bs+\bG_1\ba\|^2+\gamma\|\bG_2\ba\|^2
\end{equation}
Next, noting that $\ba$ is real-valued, but $\bs$, $\bG_1$, and $\bG_2$ are complex-valued, one may rearrange (\ref{eqn:xi2}) as
\begin{equation}\label{eqn:xi3}
\xi=\left\|\left[\begin{array}{c}
\tilde\bs\\
{\bf 0}\end{array}\right]
+\left[\begin{array}{c}
\tilde\bG_1\\
\sqrt\gamma\tilde\bG_2\end{array}\right]\ba
\right\|^2
\end{equation}
where
$$\tilde\bs=\left[\begin{array}{c}
\Re[\bs]\\
\Im[\bs]\end{array}\right],~~
\tilde\bG_1=\left[\begin{array}{c}
\Re[\bG_1]\\
\Im[\bG_1]\end{array}\right],~~
\tilde\bG_2=\left[\begin{array}{c}
\Re[\bG_2]\\
\Im[\bG_2]\end{array}\right],
$$
$\Re[\cdot]$ and $\Im[\cdot]$ denote the real and imaginary parts, respectively, and $\bf 0$ indicate a zero column vector.

Minimization of $\xi$ is a least squares problem. It has the solution
\begin{equation}\label{eqn:bao}
\ba=-\bB\tilde\bs
\end{equation}
where
\begin{equation}
\bB=\left(\tilde\bG_1^{\rm T}\tilde\bG_1+\gamma\tilde\bG_2^{\rm T}\tilde\bG_2\right)^{-1}\tilde\bG_1^{\rm T}.
\end{equation}
We also note that $\bB$ is a fixed matrix that depends on the prototype filter $g(k)$ and parameters $M$, $K_{\rm e.burst}$ and $\gamma$. The matrix $\bB$  thus can be pre-calculated and used in (\ref{eqn:bao}) to find $\ba$ for each data packet.

Energy of $s_{\rm vs.short}(k)$ beyond $K_{\rm e.burst}$ will be very small, after canceling the tail, if $K_{\rm e.burst}$ and $\gamma$ are selected properly. However, the tail will not be exactly zero due to the balance that we are making between the two objectives (\ref{eqn:s1}) and (\ref{eqn:s2}). Regarding the residual tail, either one of the following two approaches could be applied: (1) Truncate the residual tail beyond $K_{\rm e.burst}$. This could raise the out-of-band spectrum and incur interference to the edge data symbols. (2) Keep the residual tail and allow overlapping of the successive packets. Such overlap, clearly, leads to inter-packet interference. However, our numerical study reveals that such interference, when the design parameters are selected correctly, remains negligible.

\section{Receiver Design Considerations with The Proposed Tail-Shortening}\label{sec_considerations}
\begin{figure}
\centering
\includegraphics[scale=1.0]{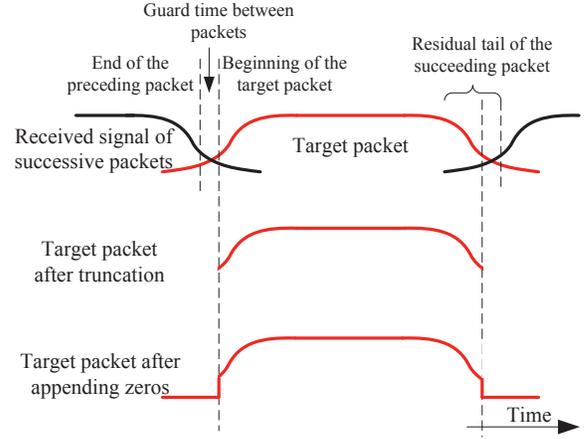}
\caption{Receiver design with overlapping of residual tails and adjacent packets (that may originate from different transmitters). }\label{fig_receiver_design}
\end{figure}

Successive packets with the proposed tail-shortening are allowed to be placed much closer in time than those without tail-shortening. Therefore, receiver with tail-shortening has to be carefully designed to avoid interference from adjacent packets. Such a receiver could work as shown by Fig. \ref{fig_receiver_design}. First, the receiver detects the beginning and end of the target packet, say, $K_{\rm b.burst}$ and $K_{\rm e.burst}$. This could be accomplished by timing synchronization mechanism of the FBMC system, which relies on the pilot/preamble symbols. Secondly, the receiver extracts the target packet from the received signal by truncating the signal at the beginning and end. Thirdly, the receiver appends certain length of zeros at both sides of the truncated packet and feeds it to a standard FBMC receiver.

If the transmitter truncates the residual tail beyond $K_{\rm b.burst}$ or $K_{\rm e.burst}$, there will be no interference from adjacent packets, as long as the guard time between packets is long enough to absorb the channel spread. If the residual tail is not  truncated, it is an interference to the packet it overlaps with. We will simulate the second case to see how the error performance is affected by the residual tail in Section \ref{sec_perf}.

\section{Performance Evaluation}\label{sec_perf}
In this section, the performance of the proposed tail-shortening method is evaluated via simulations. We simulate an FBMC system with a maximum number of subcarriers of $M=256$ which only a subset ($\Omega$) of them will be active. % from which  the number of active subcarriers  is $200$ or $12$.
We employ the PHYDYAS filter \cite{M.G.Bellanger, S.Mirabbasi, A.Viholainen} and IOTA filter \cite{Farhang-Boroujeny2011} with length $L_{\rm filter}=4M+1$. This corresponds to an overlapping factor of $\eta=4$. End of the packet is $K_{\rm e.burst}=K_{\rm f.symbol}+M/2$, where $K_{\rm f.symbol}$ is the time corresponding to the center of the final real-valued symbol (the $(N-1)$th real-valued symbol). Six virtual symbols ($V=6$) on each end of the burst are employed for the proposed tail-shortening method. Simulation results of hard truncation and truncation with windowing  are also presented for comparison. A raised-cosine window with roll-off length $L_{\rm ro}$ is applied for truncation with windowing.

To define the tail overhead of FBMC-OQAM signals, let us consider a reference OFDM signal with the same number of complex symbol ($N/2$) and without CP. The signal has a length of $NM/2$, or it ends at $K_{\rm ref}=K_{\rm f.symbol}+M/4$. We define the tail overhead of FBMC-OQAM signal with respect to  $K_{\rm ref}$ as $L_{\rm oh}=K_{\rm e.burst}-K_{\rm ref}$. In the simulations, $K_{\rm e.burst}$ is set to $K_{\rm f.symbol}+M/2$, which corresponds to an overhead $L_{\rm oh}=M/4$ or $T/4$ in time on each end.

Each FBMC burst consists of $N=14$ real-valued PAM symbols ($7$ complex OQAM symbols), where the tails on both ends add up to an overhead of $100\times (2\times T/4)/(7\times T)=7.14\%$ of the duration of the $7$ complex symbols. We selected the burst size to match the smallest resource unit in the LTE-OFDM standard \cite{dahlman20134g}. For this choice, the FBMC overhead closely matches the CP overhead of LTE-OFDM which is $7.03\%$ when a minimum length CP (often referred to as normal CP) is applied. We note that in OFDM this overhead  increases  to 25\% if an extended CP, \cite{dahlman20134g}, is used. On the other hand, in FBMC the above overhead decreases in longer data packets.

\subsection{Residual Tail and Cancelation Signal Energy}
\begin{figure}
\centering
\includegraphics[width=3.4in]{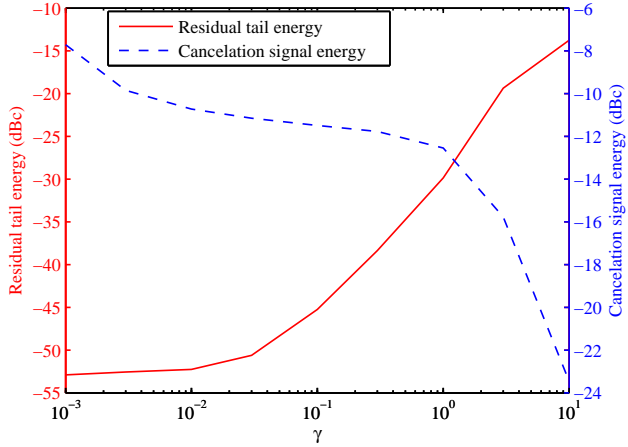}
\caption{Residual tail and cancelation signal energy, normalized by the energy of one real-valued symbol in the FBMC burst.}\label{fig_residualCancelation}
\end{figure}

Fig. \ref{fig_residualCancelation} shows the average residual tail and cancelation signal energy ($\xi_1$ and $\xi_2$, respectively) for a range of $\gamma$. Here, we have used  PHYDAYS filter and the number of active subcarriers is equal to $200$. The residual tail and cancelation signal energy are normalized with respect to the energy of each real-valued symbol (equivalently, the carrier) in the FBMC burst; the dBc unit in Fig. \ref{fig_residualCancelation} refers to decibels relative to carrier.  As observed, there is a wide range of the  parameter $\gamma$ that leads to a  well-cancelled tail while the cancellation signal energy remains small. For instance, at $\gamma=0.1$ the residual tail energy drops below $-45$~dBc, and at $\gamma=0.005$ the cancelation signal energy is still below $-10$~dBc.

Other points that worth noting (but not shown in Fig. \ref{fig_residualCancelation}) are: (i) When no tail-shortening is applied (equivalently, $\gamma\rightarrow\infty$), the average tail energy is $-12.1$dBc. (ii) When $\gamma=0$, $\xi_1=-53.3$~dBc and $\xi_2=28.0$~dBc. This value of $\xi_2$, in particular, indicates the significance of factoring the cancelation signal energy in our design equations.

\subsection{EVM Performance with Residual Tail Truncation}
As discussed before, truncation of $s(k)$ beyond  $k=K_{\rm e.burst}$ incurs distortion to the edge data symbols. This distortion is quantified by looking at the error vector magnitude (EVM) of data symbols at the beginning and the end (i.e., the edge symbols) of each burst. EVM is defined as the mean square of the errors in the detected symbols at the receiver output, normalized to the energy of each real-valued symbol, when channel is free of noise.

\begin{figure}
\centering
\includegraphics[width=3.4in]{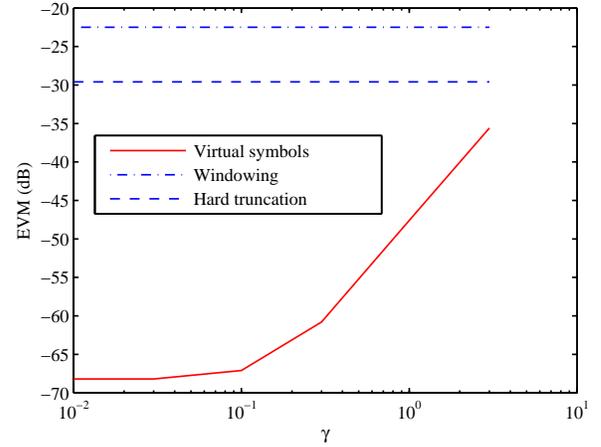}
\caption{EVM of the Edge Symbols.}\label{fig_evmPlot}
\end{figure}

Fig. \ref{fig_evmPlot} plots EVM of the edge symbols for the proposed method within  a range of $\gamma$, for the case of PHYDAYS filter and $200$ active subcarriers. Results of the hard truncation and truncation with windowing are also presented for comparison. The tails are truncated at $K_{\rm e.burst}$ for all the methods, therefore they all have the same tail overhead. The roll-off length $L_{\rm ro}$ is set to $M/4$ for the truncation with windowing method. As results show, the proposed tail-shortening by virtual symbols significantly outperforms the hard truncation and truncation with windowing. It is also interesting to note that, with respect to EVM, hard truncation outperforms truncation with windowing. This is because, for a set value of $K_{\rm e.burst}$, windowing, in addition to hard truncation, adds some distortion to the first and last $L_{\rm ro}$ samples of each burst.

\subsection{Error Performance without Residual Tail Truncation}
\begin{figure}
\centering
\includegraphics[width=3.4in]{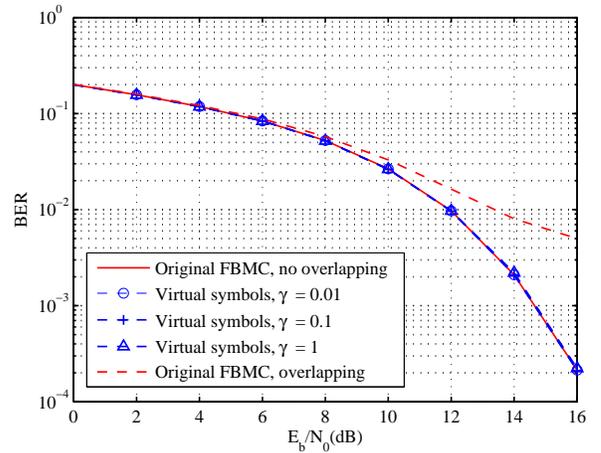}
\caption{BER performance without residual tail truncation.}\label{fig_ber_overlap}
\end{figure}

In this subsection, we simulate a transmitter that overlaps the residual tails within the adjacent packets, as shown in Fig. \ref{fig_receiver_design}. This simulation aims to quantify how BER performance is affected by overlapping the residual tails with adjacent packets. The channel under simulation is an AWGN channel and no guard time between packets is provided. Fig. \ref{fig_ber_overlap} shows the results with different values of $\gamma$, $200$ active subcarriers, $64$-QAM modulation and PHYDAYS filter. Performance of the original FBMC-OQAM system without overlapping is also provided for comparison. As observed, overlapping the residual tails with adjacent packets causes no observable influence on the BER performance, with the selected $\gamma$'s. In comparison, as also shown in Fig. \ref{fig_ber_overlap}, overlapped FBMC-OQAM packets without tail-shortening result in a significant degradation in BER.

\subsection{PAPR}
\begin{figure}
\centering
\includegraphics[width=3.4in]{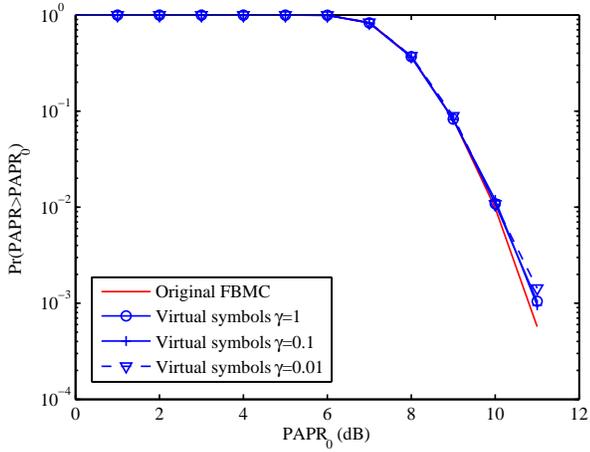}
\caption{PAPR of the FBMC bursts with the proposed tail-shortening.}\label{fig_papr_burst}
\end{figure}

Fig. \ref{fig_papr_burst} demonstrates the PAPR of the FBMC bursts with the proposed tail-shortening, for the case of PHYDAYS filter and $200$ active subcarriers. PAPR of original FBMC signal is also presented for comparison. As observed, the proposed method has a minimal effect on PAPR.

\subsection{PSD Performance}

\begin{figure}
\centering
\includegraphics[width=3.7in]{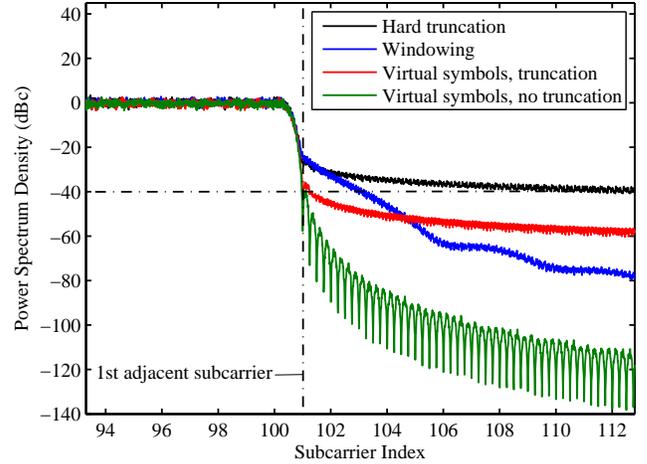}
\caption{PSD Performance of the proposed method with PHYDYAS filter, 200 active subcarrier,  $L_{\rm oh}=M/4$ and $\gamma=0.1$.}
\label{fig_psd_0.1_200sc}
\end{figure}

\begin{figure}
\centering
\includegraphics[width=3.7in]{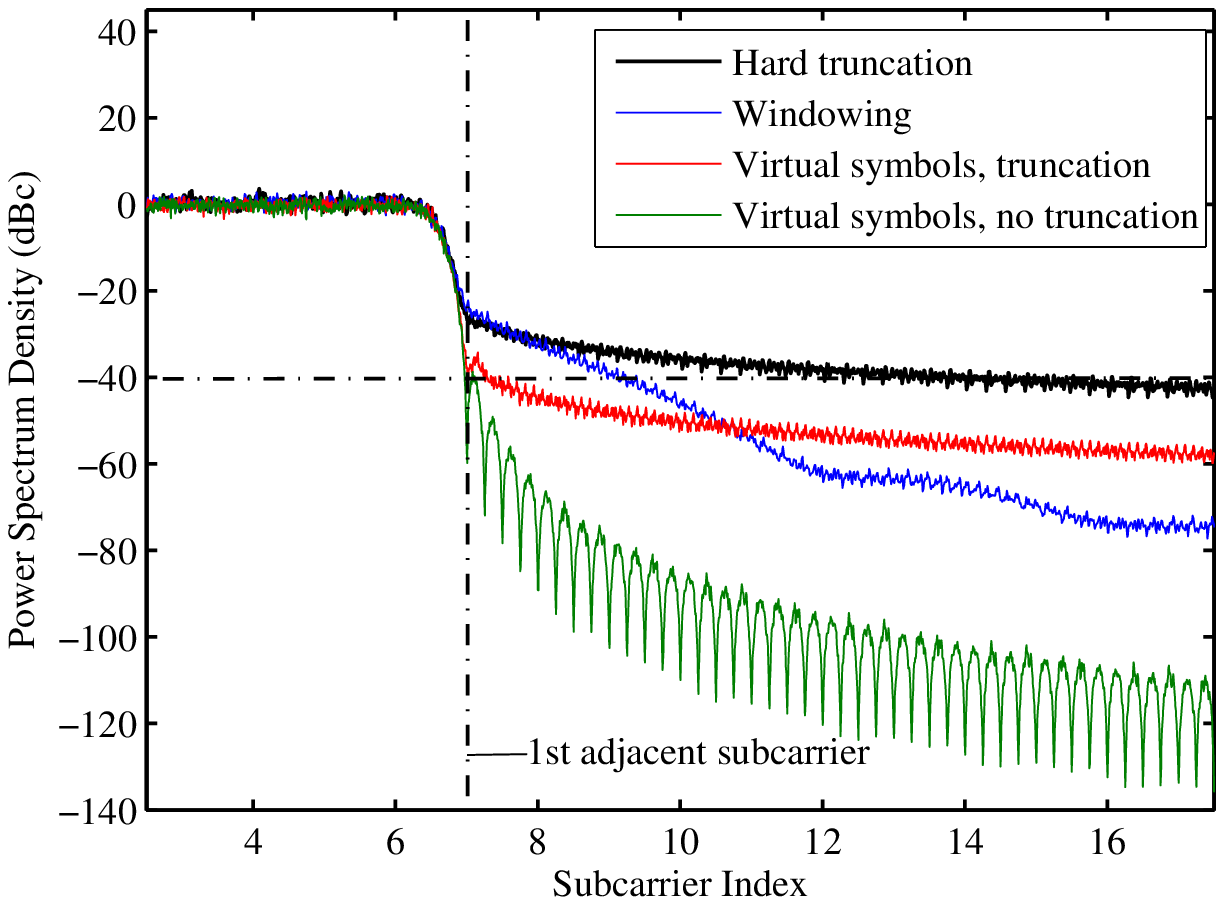}
\caption{PSD Performance of the proposed method with PHYDYAS filter, 12 active subcarrier, $L_{\rm oh}=M/4$ and $\gamma=0.1$.}
\label{fig_psd_0.1_12sc}
\end{figure}

Figs. \ref{fig_psd_0.1_200sc} and  \ref{fig_psd_0.1_12sc} show the power spectral density (PSD) of the proposed method with $200$ and $12$ active subcarriers, respectively. Here, we have adopted PHYDYAS filter and the parameters $L_{\rm oh}=M/4$ and $\gamma=0.1$ are used. The performance of the original FBMC signal and those with hard truncation and truncation with windowing are also presented for comparison. For the latter methods, the tail overhead is the same as the method proposed in this paper. Clearly, by construction, the constructed waveforms according the method proposed in this paper have exactly the same PSD as the original FBMC signal, if the residual tail is not truncated. With tail truncation, the proposed tail-shortening method still achieves an OOB suppression of about $40$dB at the first adjacent subcarrier, which is a far better performance than those of the hard truncation and truncation with windowing.

\begin{figure}
\centering
\includegraphics[width=3.7in]{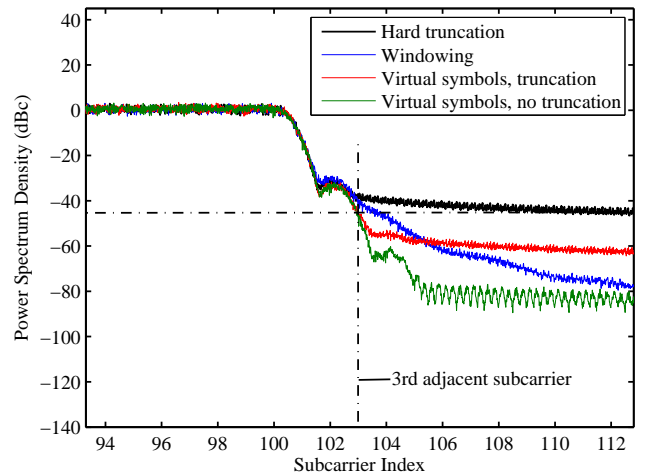}
\caption{PSD Performance of the proposed method with IOTA filter, 200 active subcarrier, $L_{\rm oh}=M/4$ and $\gamma=0.1$.}
\label{fig_psd_0.1_200sc_IOTA}
\end{figure}

For IOTA filter, the PSD of the proposed method with $\gamma=0.1$ and $200$ active subcarriers is presented in Fig. \ref{fig_psd_0.1_200sc_IOTA}. Due to better concentration in time than PHYDYAS pulse, FBMC signal with IOTA pulse has a sharper transition at the tail however higher OOB emission at adjacent subcarriers. Therefore, even with truncation, the proposed virtual symbols method has a PSD performance with an OOB emission of around $-60$~dB. Moreover, compared to the truncation with windowing, the proposed method has a deeper out-of-band suppression from the $3$rd to the $5$th adjacent subcarriers.

\subsection{Comparison with UFMC and GFDM/C-FBMC}
To complete our study and position the modified FBMC waveform against the recently proposed candidate waveforms for 5G, we compare the modified FBMC with  UFMC, GFDM, and C-FBMC waveforms in terms of their OOB emission and the packet length overhead incurred due to relevant filtering operations. Noting that GFDM and C-FBMC have a very similar OOB emission, we only present the PSD results of C-FBMC as a representative of the two.

\begin{figure}
\centering
\includegraphics[width=3.7in]{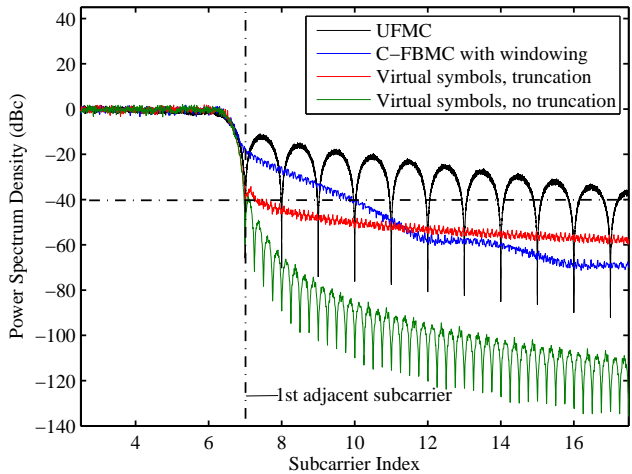}
\caption{PSD performance comparison among different multi-carrier signals, 200 active subcarrier, $L_{\rm oh}=M/4$.}
\label{fig_psd_0.1_200sc_UFMC_att40}
\end{figure}

Fig. \ref{fig_psd_0.1_200sc_UFMC_att40} presents a set of PSD plots of UFMC, C-FBMC, and the FBMC with virtual symbols. For UFMC, we have used the Dolph-Chebyshev filter with a stopband attenuation of 40 dB. In the literature, this choice has been widely advertised as the best compromised choice, \cite{wunder20145gnow,wang2015filter}. The FFT length in UFMC is set equal to 256 and the length of Dolph-Chebyshev filter is equal to 19. As seen, due to the short length of the Dolph-Chebyshev filter, UFMC has a poor OOB emission for many subcarriers before it reaches the set 40 dB attenuation. Such attenuation is achieved only at subcarriers that are more 12  subcarriers away from the last active subcarrier in the passband of the waveform. The C-FBMC has a better OOB emission performance, however, still it does not reach the 40~dB attenuation until the forth subcarrier away from the last active subcarrier in the passband. The proposed FBMC, on the other hand, easily drops below the 40~dB attenuation line at the center of the first subcarrier adjacent to the passband.  This is nearly true even after truncation of its suppressed tails.

For the results presented in  Fig. \ref{fig_psd_0.1_200sc_UFMC_att40}, in UFMC, the overhead introduced by the Dolph-Chebyshev filter is 7.03\% of the transmit signal before passing through the Dolph-Chebyshev filter.  C-FBMC and FBMC waveform parameters are chosen such both carry the same length of data (7 QAM symbols across time) and both have tail overhead of 7.14\%. For C-FBMC, all the overhead has been used to introduce raise-cosine roll-offs, meaning there is no CP in the waveform. Hence, any CP that is added to take care of the channel will increase the overhead of the C-FBMC waveform and makes it longer than that of the FBMC waveform with virtual symbols. We thus see that while the modified FBMC which is proposed in this paper offers a superior performance in terms of OOB emission, it has a tail overhead that is comparable to or lower than that of its competitors.

% the signals with $200$ active subcarriers and $L_{\rm oh}=M/4$. FBMC signals employ PHYDYAS filter in the simulation. FBMC signal with the proposed method employs $\gamma=0.1$. The C-FBMC signal employs a window of roll-off length $L_{\rm ro}=M/4$. It is observed that all these signals have satisfactory OOB performance at the $3$rd adjacent subcarrier or beyond. However, none of them is comparable with the proposed method at the first adjacent subcarrier, which is an about $40$dB OOB suppression. This outstanding OOB performance makes the proposed method ideal for the scenarios that requires high spectrum utiliza

\section{Conclusions}\label{sec_conlusion}
In this paper, we proposed a method based on virtual symbols for tail-shortening in the FBMC-OQAM waveforms. The proposed method transmits a set of virtual symbols at the two sides of each data packet to suppress signal samples at the FBMC-OQAM tails. An optimization procedure was formulated to minimize the tail energy beyond a set sample index. Extensive simulations were presented with a tail overhead of $T/4$ on each side. The results showed that while the proposed method retains the excellent OOB emission performance of the conventional FBMC method, a very low EVM, and a very minimal change in PAPR, it results in a tail overhead that remains comparable to or lower than those of the UFMC, GFDM and C-FBMC, which in recent literature have been introduced as its competitors.

\bibliographystyle{IEEEtran}

\bibliography{virtual_symbol}

\end{document}